# Collaborative Pressure Ulcer Prevention: An Automated Skin Damage and Pressure Ulcer Assessment Tool for Nursing Professionals, Patients, Family Members and Carers

Paul Fergus, Carl Chalmers and David Tully
Liverpool John Moores University, Byrom Street, Liverpool, L3 3AF, UK

**Abstract**—This paper describes the Pressure Ulcers Online Website, which is a first step solution towards a new and innovative platform for helping people to detect, understand and manage pressure ulcers. It outlines the reasons why the project has been developed and provides a central point of contact for pressure ulcer analysis and ongoing research. Using state-of-the-art technologies in convolutional neural networks and transfer learning along with end-to-end web technologies, this platform allows pressure ulcers to be analysed and findings to be reported. As the system evolves through collaborative partnerships, future versions will provide decision support functions to describe the complex characteristics of pressure ulcers along with information on wound care across multiple user boundaries. This project is therefore intended to raise awareness and support for people suffering with or providing care for pressure ulcers.

**Index Terms**—Pressure Ulcers, Classification, Deep Learning, Machine Learning, Convolutional Neural Networks.

## 1 INTRODUCTION

IN Britain, 700,000 people are affected by pressure ulcers each year [1]. Pressure Ulcers significantly impacts the quality of life of patients and costs the National Healthcare Service (NHS) approximately £2.1bn per annum to manage and treat [2]. The National Institute for Care Excellence (NICE) has provided evidence-based guidelines for pressure ulcer risk assessment and prevention [3]. The guidelines are primarily used by healthcare staff directly involved in patient care, such as district nurses, healthcare assistants and tissue viability specialists. However, despite these guidelines, managing pressure ulcers remains a significant challenge with much of the burden placed on healthcare professionals [4]. This is a practice often performed alongside the other hospital, care home and community care duties they routinely carry out.

The fastest growing subgroup of people in the UK is those over 65 years of age [5]. Combined with increased rates of obesity, diabetes, and cardiovascular disease, there is increased demand for services to assist with activities of daily living due to decreased mobility - cuts to social and GP services are likely to add to this problem as conditions are not dealt with early [6]. Reduced activity/mobility is a major risk factor in pressure ulcer development [7]. Under the current care provisions and legislation, these combined factors will negatively impact the sustainability of ongoing healthcare services [8]. Furthermore, taking into account pressure from the NHS Executive to reduce pressure ulcer prevalence, decreasing their occurrence will become much harder to achieve; however, the detection of pressure ulcers is vital for early intervention practices [9].

Pressure ulcers are considered to be a largely preventable problem and whilst the treatment and response to pressure ulcers are a predominately clinical one. The prevention of them can only be fully realised through a shared responsibility [10]. Patient and family/carer involvement should be at the core of any pressure ulcer strategy [11]. However, to date, both patients and family/carers have limited involvement, with a general lack of awareness of guidelines [12]. The literature shows that patients and family/carers would welcome a much more active role [13]. Therefore, empowering patients and involving them in the fight against pressure ulcers may help to redress the balance and have a far greater impact on pressure ulcer prevention than current intervention practices [13].

The major drawback, however, resides in raising awareness and educating patients to understand best practices and recommended guidelines, which can be difficult for healthcare professionals, let alone patients [14]. Nonetheless, guidelines do provide a comprehensive list of recommendations and should be actively lobbied among different user groups. Evidence shows that not doing so leads to inconsistency and variation in knowledge and practice, and a general feeling of uncertainty among guideline users [15]. The Pressure Ulcers Online website (http://www.pressureulcersonline.co.uk) aims to address these issues with our clinical partner Mersey Care NHS Foundation Trust by investigating the use of a non-intrusive, low cost, technological solution to formally assess and doc-

---

- *Paul Fergus, Carl Chalmers and David Tully are with the Faculty of Engineering and Technology Liverpool John Moores University, Byrom Street, Liverpool, L3 3AF, UK. (E-mail: p.fergus@ljmu.ac.uk, c.chalmers@ljmu.ac.uk, and d.a.tully@ljmu.ac.uk)*

ument ongoing pressure ulcer management in common risk areas. These areas include the buttocks, heels, hips, elbows, ankles, shoulder's spine and the back of the head.

## 2 PRESSURE ULCERS ONLIINE

The Pressure Ulcers Online platform operates as an open-access website that allows users to upload and analyse pressure ulcers. The system is fully supported on smart phones and tablets. Images of pressure ulcers can be uploaded via the website or analysed using video streams from a webcam or a camera on a mobile device with an app that can be downloaded and installed (currently only available on Android but will be rolled out to iOS). The platform is built on state-of-the-art in real-time object detection that allows images and videos to be used interchangeably. The system is non-invasive (does not require contact with skin) or obtrusive (requires no user interaction beyond the normal running of apps on a smart phone). The website will operate as a collaborative tool for pressure ulcer analysis by healthcare professionals, patients or family/carer, including those with an interest in pressure ulcer research. This is the first ever interactive machine learning website for pressure ulcer analysis that combines machine learning algorithms, in particular, convolutional neural networks for deep learning image processing and advanced high-performance server-side web processing, to assess pressure ulcer incidences. The current version will evolve and hopefully help provide a standardised platform for pressure ulcer management, training and documentation.

### 2.1 Data Collection

The system is bootstrapped with initial data collected from Google Images to train the machine learning models implemented in the system. The ethos, however, is to develop a crowd sourced solution and a continually growing central repository for pressure ulcer images and classification results. The dataset will be versioned and made available through formal data requests. Interested parties will be asked to provide basic user and organisational information and a brief description of what the data will be used for. The data provided by contributors will be utilised to further train and update our system models under the clinical guidance of Mersey Care NHS Foundation Trust. Subsequent models will include a much more clinically focused and feature rich representation of pressure ulcer characteristics and their corresponding care and prevention management plans.

### 2.2 Convolutional Neural Network

Convolutional neural networks (CNN) for deep machine learning provide new possibilities for a variety of emerging applications that were impossible until recently [16]. Since the release of Google's TensorFlow Application Programming Interface (API) in 2017 [17], computer technology for vision and image processing can now be used to detect instances of objects by class (such as humans, buildings, cars and pressure ulcers) in digital images and videos [18]. Using this technology on a website or mobile device the pressure ulcers online platform can currently capture the four main stages of pressure ulcer progression (I, II, III, IV) and the seven tissue types commonly present at the wound site (necrotic, sloughy, healthy granulating, unhealthy granulating, hyper granulating, infected and epithelizing). As the system evolves the goal is to provide a clinically relevant technology to identify the early onset of poor skin health and facilitate Early Intervention Practices (EIP). Through the user of cognitive APIs advice and guidance will be provided to aid in the treatment and management of pressure ulcers [19].

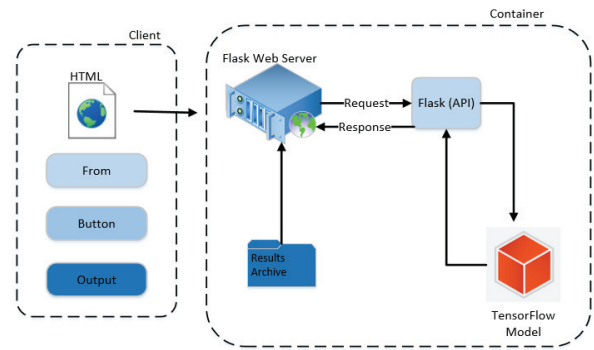

Fig. 1: TensorFlow Serving

CNNs are deep feed-forward artificial neural networks commonly used to analyse images and can be viewed as a variation of a multilayer perceptron. They were inspired by biological connectivity patterns where individual cortical neurons respond to stimuli in the receptive field [20]. CNNs use minimal pre-processing when compared with other image classification algorithms, given that the network learns its own filters unlike traditional algorithms where they were hand-engineered, which is a major advantage of CNNs. A basic CNN structure comprises convolutional layers, pooling layers and a fully connected layer. Other layers such as dropout and early stopping layers are also used for regularisation.

In our system, a CNN architecture is combined with transfer learning (a model developed for a specific task is reused as the starting point for our pressure ulcer model). This is a common approach in deep learning when vast compute, and time resources are required. More importantly, it is also a popular approach when access to large image datasets are possible. For the initial phase of the project only limited data was available. Therefore, the transfer learning strategy was necessary.

### 2.3 End-to-End Web Container

In order to host and serve the TensorFlow model a web server running a Flask REST API and associated Flask web server has been deployed. The server hosts a python distribution containing all the dependencies, which include TensorFlow 1.6 (CPU), Flask 1.0.2 and Flask WTF 0.14.2. The site is hosted on Windows Server 2012R2 with a Celeron G1619T @ 2.3GHz and 4GB of RAM. The frozen inference graph, model checkpoint and label PBTXT file is hosted within the Flask distribution. Figure 1 highlights the end-to-end deployment.

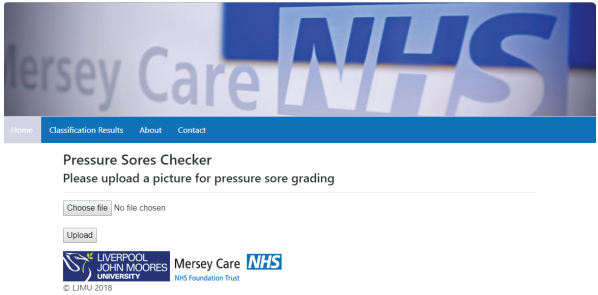

Fig. 2: TensorFlow Serving

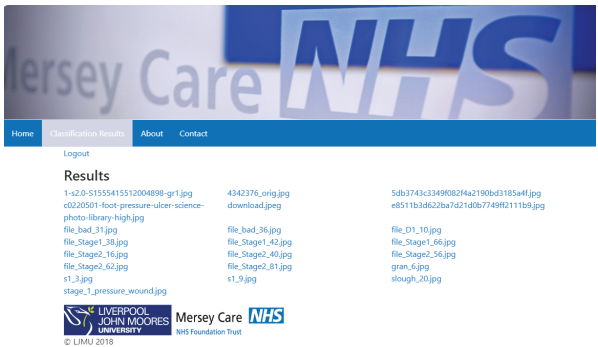

Fig. 3: TensorFlow Serving

The website enables the user to select and upload images for classification (webcam support will be added in subsequent versions). The web server forwards the request to the Flask API to return the classification results.

Figure 2 highlights the home page where images can be uploaded for classification and later verified.

Figure 3 highlights the results page, which is an archive of all the images and the associated classification.

## 3 PRELIMINARY RESULTS

The ssd_mobilenet_v1_coco_11_06_2017 model provides a base model for training. Utilising transfer learning the initial training phase is performed using an image set compiled from 700 images extracted from Google Images and graded by an NHS District Nurse (100 images for each class, Esch-Necro, Granulating, Stage-I, Stage-II, Stage-III, Stage-IV, and Slough). The overall quality of the images vary. However, it has been possible to train a convolutional neural network using transfer learning to produce reasonable results. The current version of the model was trained over 86000 epochs with a final loss value under approximately 0.8%. The results from some of the images classified are illustrated in Figures 4 through 7.

Figure 4 shows an image of a heel with a pressure ulcer with most of the inner section consisting of Slough. The system on this occasion identifies the Slough with a 51% confidence value.

Figure 5 illustrates a slightly more complex wound, and the system makes two classifications. The inner part of the wound contains Slough, and this is identified with a 70% confidence value. The wound as a whole has also been identified as a Stage-III wound with a 50% confidence value.

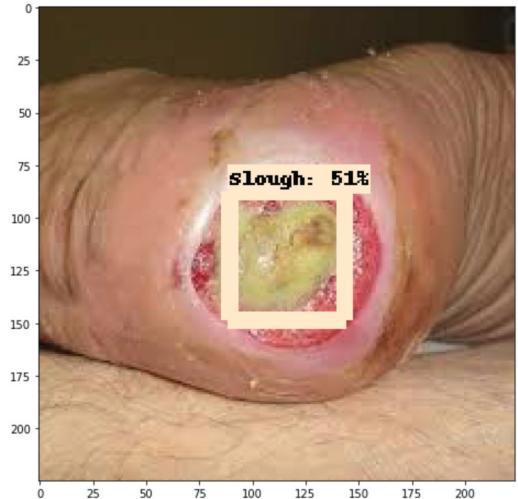

Fig. 4: Slough Classification

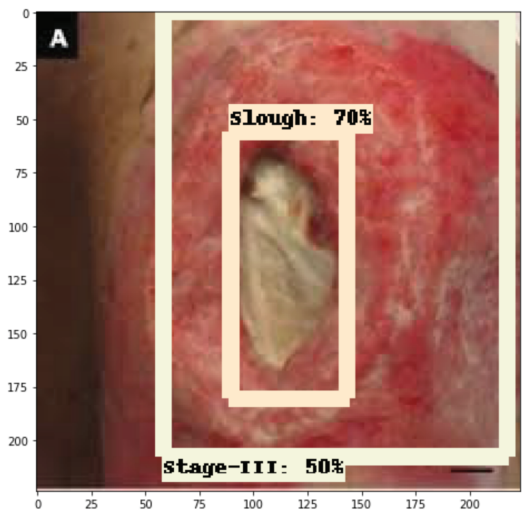

Fig. 5: Slough and Stage-III Classification

In figure 6, the wound on this occasion is clear of Slough and has been identified as a Stage-II wound with an 81% confidence value.

In the final figure (figure 7) it is less obvious whether a wound is actually present. This is typical of Stage-I pressure ulcers. The system has classified this as such with a high confidence value of 98%.

## 4 CONCLUSION

The Pressure Ulcer Online system is a first step solution towards a new and innovative platform for helping people to detect, understand and manage pressure ulcers. This position paper discusses the main problem and motivation behind the project and our aspirations to provide a central point of contact for pressure ulcer analysis and ongoing research. Using state-of-the-art technologies in convolutional neural networks and transfer learning along with end-to-end web technologies, this platform allows pressure ulcers to be analysed and findings to be reported. During the project, the system will evolve and provide decision support and inference reporting functionality to describe the complex characteristics of pressure ulcers submitted for analysis along with information on wound care across multiple user boundaries. The end-to-end website solution is the first of its kind worldwide to offer an open-access system for analysing pressure ulcers online. Our primary motivation is to raise awareness or pressure ulcers and their management and provide support to people suffering with or caring for people with pressure ulcers.

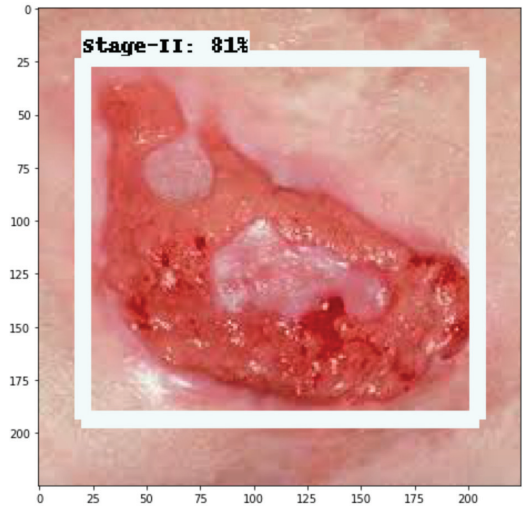

Fig. 6: Stage-II Classification

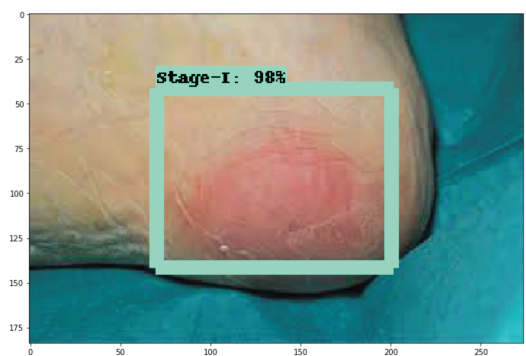

Fig. 7: Stage-I Classification

The research question under investigation is whether our proposed system can accurately detect pressure ulcer stages and tissue types and be developed into a clinical decision support system. If successful, it will provide automated documentation of pressure ulcer sites and the detailed characteristics of the wound with fully supported inference models to justify decision making points. Initial measureable outcomes would be the percentage of medical classifications and assessments performed by medical practitioners versus the classification results produced by our machine learning system. This would also be a unique opportunity to seek clinical support in the labeling of additional images to capture a much more in-depth analysis of wounds in general and to increase the accuracy of our machine learning models.

Looking further ahead we aim to solicit clinical trials with our partners to investigate remote examination and documentation in a clinical setting and the impact this might have on improving quality of life measures for patients dealing with poor skin health, including how it affects patients and healthcare professional's time, in terms of unnecessary visits, and potential cost savings. Further outcome measures will include the standardisation of pressure ulcer reporting (i.e. types of pressure ulcer, location and classification systems used). If successful, the system will correctly identify pressure ulcers, staging and automatically generate detailed and accurate documentation on the size of the wound, including its location, surrounding skin condition, and presence of tissue undermining and tunnelling, without patient or family/carer intervention beyond simply taking a photo or short video of the wound(s). Combined with clinical data, future machine learning models could be further used to personalise treatments based on patient-specific factors such as age, skin wrinkling, shearing force risk and routine diet and exercise regimes. In order to obtain this level of analysis, we are currently investigating the use of capsule networks to support this level of granularity [21].

## ACKNOWLEDGEMENT

The authors would like to thank Lorna Bracegirdle, who is a caseload holder working for Mersey Care NHS Foundation Trust for sharing her valuable knowledge on pressure ulcers and for grading the pressure ulcer images used to train our models. The authors would also like to thank Michelle Gallagher and Lesley Newport from the Skin Care Service at Mersey Care NHS Foundation Trust for evaluating the initial system. We would also like to thank Rachel Hastie from NHS Aintree Park Group Practice for reviewing our work and providing valuable feedback on the working system.

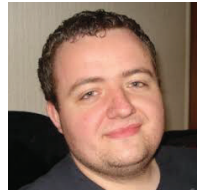

**Dr Carl Chalmers** is a Senior Lecturer in the Department of Computer Science at Liverpool John Moores University. Dr Chalmers's main research interests include the advanced metering infrastructure, smart technologies, ambient assistive living, machine learning, high performance computing, cloud computing and data visualisation. His current research area focuses on remote patient monitoring and ICT-based healthcare. He is currently leading a three-year project on smart energy data and dementia in collaboration with Mersey Care NHS Trust. As part of the project a six month patient trial is underway within the NHS with future trials planned. The current trail involves monitoring and modelling the behaviour of dementia patients to facilitate safe independent living. In addition he is also working in the area of high performance computing and cloud computing to support and improve existing machine learning approaches, while facilitating application integration.

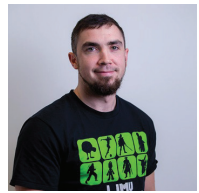

**Dr David Tully** is a Senior Lecturer in Computer Games Technology and Software Development. Dr Tully's main area of research is the use of Game Technologies for the visualisation of Big-Data and real-world data processing, including real-time rendering. In particular creating real-world city scapes where data can be combined, classified, categorised, and maniputaled in real time for treaching and training tasks including better understanding big data.

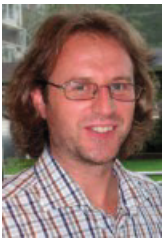

**Dr Paul Fergus** is a Reader (Associate Professor) in Machine Learning. Dr Fergus's main research interests include machine learning for detecting and predicting preterm births. He is also interested in the detection of foetal hypoxia, electroencephalogram seizure classification and bioinformatics (polygenetic obesity, Type II diabetes and multiple sclerosis). He is also currently conducting research with Mersey Care NHS Foundation Trust looking at the use of smart meters to detect activities of daily living in people living alone with Dementia by monitoring the use of home appliances to model habitual behaviours for early intervention practices and safe independent living at home. He has competitively won external grants to support his research from HEFCE, Royal Academy of Engineering, Innovate UK, Knowledge Transfer Partnership, North West Regional Innovation Fund and Bupa. He has published over 200 peer-reviewed papers in these areas.